 \documentclass[final,3p,times,twocolumn]{elsarticle}
\usepackage{amssymb}
\journal{Physics Letters B}

\usepackage{amsmath}
\usepackage{slashed}
\usepackage{amssymb}
\usepackage{braket}
\usepackage{graphicx}
\usepackage{amsmath}
\usepackage{braket}
\usepackage{epigraph}
\usepackage{tensor}
\usepackage{enumerate}
\usepackage{subfig}
\usepackage{verbatim}
\usepackage{pstricks}
\usepackage{epstopdf}
\usepackage{axodraw4j}
\usepackage{tikz}
\usepackage{pgfplots}
\pgfplotsset{/pgf/number format/use comma,compat=newest}
\usepackage{color}

\usepackage{feynmp}
\makeatletter
\def\endfmffile{%
  \fmfcmd{\p@rcent\space the end.^^J%
          end.^^J%
          endinput;}%
  \if@fmfio
    \immediate\closeout\@outfmf
  \fi
  \ifnum\pdfshellescape=\@ne
    \immediate\write18{mpost \thefmffile}%
  \fi}
\makeatother

\begin{document}

\begin{frontmatter}


%\title{A gravitational mechanism for C violation and baryogenesis.\tnoteref{label1}}
%\tnotetext[label1]{}
\author{J. I. McDonald}
\ead{pymcdonald@swansea.ac.uk}
\author{G. M. Shore}
\ead{g.m.shore@swansea.ac.uk}
\address{Department of Physics, College of Science, \\
Swansea University, Singleton Park, Swansea, SA2 8PP.}

%
%\author{J. I. McDonald\corref{cor1} and G. M. Shore\corref{cor2} }
%\ead{pymcdonald@swansea.ac.uk, g.m.shore@swansea.ac.uk}
%\address{Department of Physics, College of Science, \\
%Swansea University, Singleton Park, Swansea, SA2 8PP. }


\title{Radiatively-induced gravitational leptogenesis}


\author{}

\address{}

\begin{abstract}

We demonstrate how loop effects in gravitational backgrounds lead to a
difference in the propagation of matter and antimatter, and show 
this is
forbidden in flat space due to CPT and translation invariance. This
mechanism, which is naturally present in beyond the standard model
(BSM) theories exhibiting C and CP violation, generates a curvature-dependent chemical potential for leptons in the low-energy effective Lagrangian, allowing a matter-antimatter asymmetry to be generated in thermodynamic equilibrium, below the BSM scale.

\end{abstract}

\end{frontmatter}



\section{Introduction}

The origin of the matter-antimatter asymmetry of the universe remains one of the outstanding questions in particle physics and cosmology.
Following the framework of the celebrated Sakharov \cite{Sakharov}
conditions, a popular and long-standing explanation has involved the
out-of-equilibrium decay of heavy particles \cite{FukYan,KolbTurner},
in which matter and antimatter are produced at different rates due to
C and CP violation in the underlying theory.  An alternative to this
picture was proposed by Cohen and Kaplan \cite{Cohen} who noted that
an asymmetry could, in fact, be generated \textit{in equilibrium}
through the coupling of C and CP violating operators
involving the baryon or lepton currents to background fields,
\textit{e.g.}, $\partial_\mu \Phi j^\mu$ for a background scalar field. For
isotropic background fields, this results in a chemical potential
proportional to the time derivative $\dot{\Phi}$. More recently,
Davoudiasl \textit{et al.} \cite{Davoudiasl:2004gf} built on this idea
by suggesting gravity could play the same role as $\Phi$ with an
interaction $\partial_\mu R j^\mu$, where $R$ is the Ricci
scalar. Since then, many authors have gone on to postulate
gravitational couplings as a means of generating matter asymmetry
\cite{Alexander:2004us,
  Lambiase:2006md,Lambiase:2011by, Lambiase:2013haa,
  Ellis:2013gca,McDonald:2014,Cesare:2014,Lambiase:2015}. However, with the exception of \cite{Alexander:2004us} (where the gravitational coupling arises from the axial anomaly), in almost all of these papers the required operators are introduced by hand, with no account of their dynamical origin, in the expectation that they may arise from some unspecified, more fundamental theory.

In this Letter, we present a new mechanism for gravitational
leptogenesis in which the matter-antimatter asymmetry is generated
dynamically at the quantum loop level, without the need to postulate
additional interactions beyond the minimally coupled Lagrangian.
Specifically, we show how in a C and CP violating theory, in which the light leptons are coupled to heavy states with mass $M$,
the effective Lagrangian describing low-energy physics below this
scale involves operators coupling directly to the background
curvature, including the C and CP violating interaction $\partial_\mu R j^\mu/M^2$, which
leads to a lepton-antilepton asymmetry.\footnote{ A careful analysis
  of the modification to the 
dispersion relations implied by an operator $\partial_\mu R j^\mu/M^2$
has been given 
recently in \cite{McDonald:2014}, showing the same implications for
lepton-antilepton 
asymmetry as follow from the interpretation of $\dot{R}$ as a chemical
potential \cite{Cohen,Davoudiasl:2004gf}.}
\footnote{Another way to motivate the appearance of matter-antimatter
  asymmetry is to view $\partial_\mu R \sim \dot{R}$ as a fixed
  background coupling to the CPT odd current $j^\mu$. In this sense,
  as originally presented in \cite{Cohen}, the effect can be thought
  of as an ``environmental CPT violation'', with phenomenological
  consequences normally associated with a genuine breaking of CPT
  symmetry. The full operator $\partial_\mu R j^\mu$ is however CPT
  invariant. See \cite{McDonald:2014} for a further discussion.}
The coupling of C and CP
violating operators to a time-dependent gravitational field
circumvents the third Sakharov condition and allows the 
lepton-antilepton asymmetry to be generated in equilibrium.

The presence of explicit curvature-dependent operators in the effective Lagrangian represents a violation of the strong equivalence principle
\cite{Drummond:1979pp, Shore:2004sh}.
The physical picture is that, at loop level, the light leptons
propagate surrounded by a self-energy cloud of virtual particles,
including the heavy states. This virtual cloud has a length scale of
order $1/M$ and so interacts with the background gravitational field through tidal, curvature-dependent
forces, while its composition encodes the dynamics and symmetries of the heavy particles. In this way, gravity probes the physics of the high-scale fundamental theory and transmits this information to the low-energy effective Lagrangian describing the light leptons.

The existence of C and CP violating operators coupling to the
curvature leads directly to a difference in the propagation of matter and antimatter. 
Of course, this would be inconsistent with the (strong) equivalence principle and, in particular, could not occur in flat space.
In sec. \ref{propagationandCPT}, we give a formal proof that CPT and
translation invariance forbids this situation for interacting
theories in Minkowski space, regardless of whether there is any C (or CP)
violation in the theory, which is of course a \textit{necessary}
condition for asymmetric propagation. Conversely, when gravity, C and
CP violation are present, we show there is indeed a difference in the
propagation of matter and antimatter. Only if all these conditions are
met will this happen, meaning that the effect is intrinsic to
gravitational backgrounds and not simply a consequence of C and CP violation already present in the original Lagrangian.   

The mechanism described here is very general. For clarity, however, we illustrate it in a specific model familiar in the BSM literature,
namely the ``see-saw'' Lagrangian, in which the light, left-handed lepton doublets
$\ell_i$ ($i = e, \mu , \tau$) and Higgs field\footnote{In this
  notation, the Higgs doublet $\tilde{\phi}$ appearing in the SM lepton
  sector is related by $\phi^a = \epsilon^{ab} \tilde{\phi}^{\dagger
    b}$.} are coupled to heavy right-handed sterile neutrinos
$N_\alpha$ with non-degenerate masses $M_\alpha$ ($\alpha=1,\ldots n$):
\begin{align}
\mathcal{L} =\sqrt{-g}\left[ \overline{N} \slashed{D} N+ \lambda_{i \alpha} \bar{\ell}_i\phi  N_\alpha   + \frac{1}{2}\overline{\left(N^c\right)} 
\, M \,  N  + \mbox{h.c.}\right]  \ .
\label{FYmodel}
\end{align} 
$\lambda_{i\alpha}$ is a complex Yukawa matrix, providing the required
C and CP violation.  For clarity, we omit any explicit labelling of L and R
handed fields in what follows. 

This is simply the model used by Fukugita and Yanagida \cite{FukYan} in their original demonstration of leptogenesis in flat space through the out-of-equilibrium decays of the heavy neutrinos,  $N_\alpha \rightarrow  \ell_i \phi^*$. Rather than using the heavy neutrinos in this way, however, 
we integrate them out to obtain the low-energy effective Lagrangian describing the physics of the leptons $\ell_i$ below the BSM scale $M_\alpha$.
In this Letter, we show that this gives rise to the operator
\begin{equation}\label{Operator}
\mathcal{L}_i = \partial_\mu R \, \bar{\ell}_i \gamma^\mu \ell_i \,
\sum_{\alpha,  \,  \beta, \,   j} \frac{\mbox{Im}\left[
    \lambda^\dagger_{\beta i} \lambda_{i\alpha} \lambda^\dagger_{\beta
      j} \lambda_{j \alpha} \right]}{3 M_\alpha M_\beta} I_{[\alpha \beta] }
\end{equation}
The function $I_{\alpha \beta}= I\left (M_\alpha,M_\beta\right)$, which we calculate exactly, gives an antisymmetric part under interchange of $M_\alpha$ and $M_\beta$ 
and is determined from a certain class of two-loop self-energy diagrams.
Eq.(\ref{Operator}) is of exactly the form required to generate a
lepton-antilepton asymmetry. This is maintained by $\Delta L=2, \,$
$\phi \ell^c \leftrightarrow \phi^* \ell$ reactions in equilibrium which, in the
conventional heavy-decay model, can wash out the lepton asymmetry
but are essential in our scenario.
We therefore have a mechanism for radiatively-induced gravitational leptogenesis, in which the asymmetry can be generated in equilibrium long after the decay of the heavy particles, at energies and temperatures well below their mass.

\section{Propagation and CPT}\label{propagationandCPT}

In a C invariant theory, the propagation of matter and anti-matter will be identical, so the presence of complex $\lambda$ is crucial to have an asymmetry in matter/anti-matter propagation, regardless of the background. We now show that the propagation of matter and antimatter must be the same in any theory in which translation and CPT symmetry holds. We demonstrate this explictly for spin 1/2 Dirac fermions. CPT symmetry is realised by an anti-unitary operator $\Theta$ such that the lepton propagator satisfies
\begin{align}
S_{ab}(x',x) & =\braket{ \ell_a(x') \bar{\ell}_b (x)} \nonumber \\
& = \braket{ \left(\Theta \ell_a(x') \Theta^{-1} \right) \left( \Theta
    \bar{\ell}_b (x) \Theta^{-1}\right)}^*  \ ,
\nonumber \\ 
\end{align}
where $a,b$ label spinor components. The CPT transformations can be written as $\Theta \ell (x') \Theta^{-1} = \gamma^0\gamma_5 C^{-1} \ell^c(-x)$ and $\Theta \bar{\ell}(x') \Theta^{-1} =\overline{\ell^c} (-x')C\gamma_5 \gamma^0$, where $\ell^c = C \bar{\ell}^T$ is the Dirac charge conjugate and $C$ is the charge-conjugation matrix satisfying $C \left(\gamma^\mu\right)^T C^{-1} =- \gamma^\mu $. Inserting these expressions, and taking note of the overall complex conjugation, we find, after some algebra
\begin{equation}\label{SSc}
S (x',x) = \gamma_5 C [S^c(-x,-x')]^T C^{-1}\gamma_5 \ ,
\end{equation}
where $S^c(x,y) = \braket{\ell^c(x) \overline{\ell^c}(y)}$ is the antiparticle propagator. Translation symmetry means that $S^c(x,y) = S^c(x-y)$ which implies that $S^c(-x,-x') = S^c(x',x)$. From Lorentz invariance (inherent to a discusssion of spinors) we can write
\begin{equation}
S^c(x',x)  = S^c(x'-x)= \int \! \frac{d^d p}{(2\pi)^d} \left[A(p^2)
  \slashed{p}  + B(p^2)\right]e^{-i p \cdot(x'-x)} 
\end{equation}
for some functions $A$ and $B$. Substituting this expression into (\ref{SSc}) and using the properties of the matrix $C$ gives
\begin{equation}
S (x',x)=S^c(x',x) \ ,
\end{equation} 
establishing that matter and antimatter propagate identically in a translational invariant and CPT conserving theory.

We now examine how loop corrections in gravitational backgrounds, which in general violate translation symmetry, can create a difference in lepton and antilepton self-energies $\Sigma(x,x')- \Sigma^c(x,x')$ associated to the propagators $\braket{\ell(x) \overline{\ell}(x')}$ and $\braket{\ell^c(x) \overline{\ell^c}(x')}$. 

First, note that in the model of Eq.(\ref{FYmodel}), the Majorana mass term for the heavy neutrinos means that there are two classes of propagators, \textit{charge-violating} propagators $S^\times_{\! \alpha}(x,x')  = \braket{N (x)\overline{N^c}(x')}$ and \textit{charge-conserving} propagators $S_{\! \alpha}(x,x')=\braket{N_\alpha (x)\overline{N}_\alpha(x')}$ where the $C$ script denotes the Dirac charge conjugate. In flat space, translation invariance allows us to write them in momentum space as
\begin{equation}\label{flatprops}
S_\alpha(p) = \frac{i \slashed{p}}{p^2 - M_\alpha^2}, \qquad
S^\times_\alpha(p) = \frac{i M_\alpha}{p^2 - M_\alpha^2} \ .
\end{equation}
As we see below, the charge violating propagators are key to generating a matter-antimatter asymmetry.

At one loop (see figure \ref{oneloop}), the lepton and anti-lepton propagators are the same:
\begin{figure}
\centering
\includegraphics[scale=0.38]{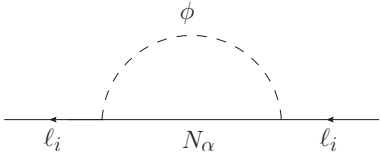}
\caption{One-loop lepton self-energy}
\label{oneloop}
\end{figure}
\begin{equation}
\Sigma_i(x,x') = \Sigma^c_i(x,x') =\sum_\alpha\lambda^\dagger_{\alpha i} \lambda^{\!}_{i\alpha}
G(x,y)S_\alpha(x,y) \ .
\end{equation}
However, at two loops there are two diagrams (figure \ref{2loop}), which give non-zero contributions to $\Sigma(x,x')-\Sigma^c(x,x')$.
\begin{figure}
\centering
\includegraphics[scale=0.45]{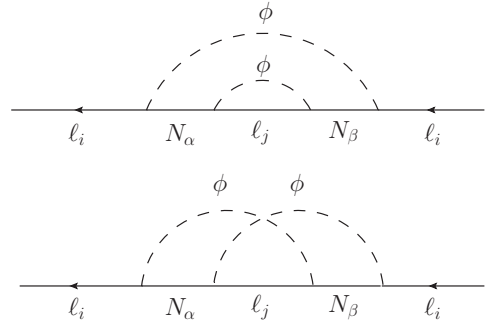}
\caption{Two-loop corrections to lepton self-energies giving non-zero contributions to $\Sigma-\Sigma^c$.}
\label{2loop}
\end{figure}
For instance, in the case of the charge violating heavy neutrino propagators, the first diagram gives
\begin{align}\label{propagator}
&\Sigma_i(x,x') - \Sigma^c_i(x,x') \nonumber \\
&=   \sum_{\alpha, \, \beta, \, j} \mbox{Im}\left[ \lambda^\dagger_{\beta i} \lambda_{i\alpha} \lambda^\dagger_{\beta j} \lambda_{j \alpha} \right] \nonumber \\
& \times G(x,x') \int d^4 y \int d^4 z \, \, G(y,z)
S^{\times}_{[\alpha} (x,y)S_j(y,z) S^{\times}_{\beta]}(z,x') \ ,
\end{align}
whilst the second gives
\begin{align}\label{vertex}
&\Sigma_i(x,x') - \Sigma^c_i(x,x') \nonumber \\
&=     \sum_{\alpha, \, \beta, \, j}  \mbox{Im}\left[ \lambda^\dagger_{\beta i} \lambda_{i\alpha} \lambda^\dagger_{\beta j}\lambda_{j \alpha} \right] \nonumber \\
& \times \int d^4 y \int d^4z  \,  G(y,x') G(x,z)  S^{\times}_{[\alpha
}(x,y)S_j(y,z) S^{\times}_{\beta]}(z,x') \ .
\end{align}
Notice that we have antisymmetrised over $\alpha$ and $\beta$ in the
integral since  $\mbox{Im}\left[ \lambda^\dagger_{\beta i}
  \lambda_{i\alpha} \lambda^\dagger_{\beta j} \lambda_{j \alpha}
\right] $ is antisymmetric in $\alpha,\beta$. For the other type of
heavy neutrino propagator, only the first diagram contributes (due to
charge considerations) and the expression is similar to (\ref{propagator}) but with a
Yukawa matrix contribution $\mbox{Im}\left[ \lambda^\dagger_{\beta i}
  \lambda_{i\alpha} \lambda^\dagger_{\alpha j} \lambda_{j \beta}
\right]$. 
It is now clear that Eqs.(\ref{propagator}) and (\ref{vertex}) are
non-vanishing in curved spacetime.
We therefore see that as a consequence of breaking translation
invariance by a general background, there is a difference in
the propagation of matter and antimatter at two loops. 

Given the general proof above, we must also find that if we restore
translation invariance by going to Minkowski space, 
$(\ref{propagator})$ and $(\ref{vertex})$ will vanish. 
Indeed, substituting the flat space propagators of (\ref{flatprops}), we
see explicitly that the integral is symmetric under interchange of $\alpha$ and $\beta$, and $\Sigma - \Sigma^c =0$ as expected.

\section{An effective action for leptons}\label{effective action}
%We now turn to the question of matter/anti-matter propagation in curved space. Computing two-loop self-energies for a general background is a more or less impossible task. There are only a handful of backgrounds for which propagators are known exactly, and even for these special cases the two loop integrals would be incalculable. 
%
%Instead we will take a different and more insightful approach and will 

We study the dynamics of leptons at the quantum loop level using an
effective action in curved spacetime, 
valid at energies below the heavy neutrino mass scale $M_\alpha$, {\it i.e.} we integrate out the heavy neutrinos.

The fundamental physics of how gravity affects the propagation of particles in curved backgrounds at loop level
is now well understood (see, {\it e.g.}  \cite{McDonald:2014,Drummond:1979pp,Ohkuwa:1980jx, Hollowood:2008kq, Hollowood:2011yh}). 
As an interacting particle propagates, it becomes surrounded by a screening cloud of virtual particles, acquiring an effective size and, as a result, experiences tidal forces from background curvature. Hence, the effective action, which captures the effect of quantum loops, will involve interactions between particle fields and background curvature. 
The fundamental Lagrangian respects the strong equivalence principle,
by virtue of minimal coupling to gravity through the connection only,
so particles and antiparticles propagate identically at tree level. However, the
interaction of the gravitational field with this virtual cloud
violates strong equivalence, causing the dynamics to become sensitive
to the background curvature at loop level. As a result, the effective
lagrangian will contain strong equivalence violating operators which
couple the curvature tensor to lepton fields, allowing -- depending on
the structure of the cloud -- the generation of C and CP violating operators such as $\partial_\mu R \bar{\ell} \gamma^\mu \ell$.

Since we are interested in the propagation of leptons, we consider an
effective action which is quadratic in the lepton field, so that tidal
effects manifest themselves as couplings between the Riemann tensor
$R_{\mu \nu \rho \sigma}$ (and its various contractions) and fermion
bilinears $\bar{\ell} (\cdots) \ell$. The most general such action,
consistent with the symmetries of the tree-level action, namely
general covariance and gauge symmetry, was discussed in detail in
\cite{McDonald:2014}. To leading order in the mass dimension of the couplings, it consists of operators of the form
\begin{align}
\mathcal{L}_{eff} = \sqrt{-g}   
\Bigg[   \bar{\ell}i \slashed{D} \ell &+ ia\bar{\ell}\left(2R_{\mu
    \nu}\gamma^\mu D^\nu   
+ \frac{1}{2}\partial_\mu R \gamma^\mu \right) \ell \nonumber\\
&+ b \partial_\mu R  \bar{\ell} \gamma^\mu \ell\nonumber\\
&+ ic \bar{\ell}\left( 2R \slashed{D}  +  \partial_\mu R \gamma^\mu
\right) \ell \nonumber\\
& + id \bar{\ell}\left( 2 D^2\slashed{D}+ \frac{1}{4}\partial_\mu R
  \gamma^\mu \right)\ell \Bigg] \ , \label{eff}
\end{align} 
where $a,b,c,d$ are real effective couplings of mass dimension minus
two, which will depend on $\lambda_{i\alpha}$ and the masses $m_H$ and 
$M_\alpha$ in the loops. There is one term in this effective action
which is of great importance for leptogenesis and is the only C and CP
violating operator in (\ref{eff}), {\it viz.}
\begin{equation}
\mathcal{L}_{CPV} = b \, \partial_\mu R  \, \bar{\ell} \gamma^\mu
\ell \label{dmuR} \ .
\end{equation}
A careful discussion of the action of C, P and T on each of the
operators appearing in $\mathcal{L}_{eff}$ is given in \cite{McDonald:2014}.

We compute the effective coupling $b$ by matching the full and effective theories. We can capitalise on the fact that the effective couplings are independent of the choice of background and work in a conformally flat metric 
\begin{equation}\label{metric}
g_{\mu \nu} = \Omega^2 \eta_{\mu \nu} = (1 + h)\eta_{\mu \nu} \ ,
\end{equation}
which is sufficient to distinguish the various components of the
effective Lagrangian (\ref{eff}).
The computation is also simplified if we work with conformally rescaled fields, 
\begin{equation}
N\rightarrow  \Omega^{-(n-1)/2}N, \quad   \ell  \rightarrow  \Omega^{-(n-1)/2}\ell,   \quad \phi \rightarrow \Omega^{-(n-2)/2}\phi .
\end{equation}
After conformal rescaling, gravity enters only via
\begin{equation}\label{Omega}
\mathcal{L}_{\Omega}  = \frac{1}{2}\Omega \overline{N^c} M N +
\Omega^2  \left(  m_H^2 - \frac{R}{6}  \right)\phi^\dagger \phi \, + \,
\Omega^{-(n-4)/2}\lambda_{i \alpha} \bar{\ell}_i \phi N \ ,
\end{equation}
where $R = - 3 \partial^2 \Omega^2$ is the Ricci scalar for (\ref{metric}). This can then be expanded to linear order in $h$ to give $\mathcal{L}_{\Omega} = h(x) \mathcal{O}(x)$. The effective couplings can be computed by matching the transition matrix elements $\bra{\,\ell(p') \,}  \mathcal{O} \ket{\, \ell(p)\,}$ to the effective amplitudes (see in particular \cite{Ohkuwa:1980jx, McDonald:2014}, as well as \cite{Drummond:1979pp, BerGas,DHGK}, for more details). Since $R = - 3 \partial^2 h$, the contribution to the effective vertex from the operator $\mathcal{L}_{CPV} = b \partial_\mu R \bar{\ell} \gamma^\mu \ell$,  gives a contribution of the form shown in figure \ref{effective}.

\begin{figure}[h!]
\centering
\includegraphics[scale=0.4]{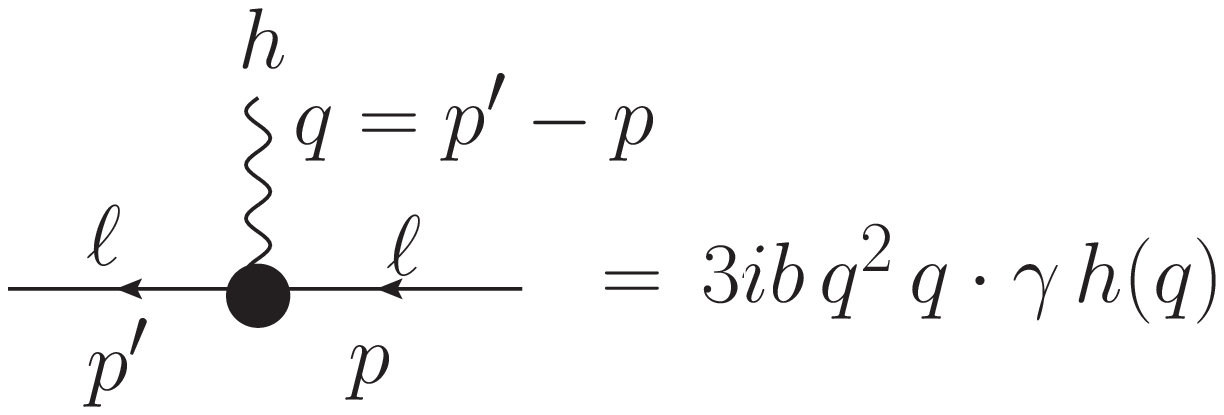}
\caption{The effective $h$ vertex, where $q=p'-p$ is the momentum transfer between the ingoing and outgoing lepton}
\end{figure}\label{effective}
For phenomenological reasons related to leptogenesis (which we will
explain below), we are only interested in diagrams which involve the
charge-violating neutrino propagators $\braket{N(x)
  \overline{N^c}(x')}$. For this kind of heavy neutrino propagator,
there are in fact no additional contributions from $h$ at the Yukawa vertex
$\mathcal{L}_\lambda =- \frac{1}{2}(n-4) h \lambda_{i \alpha}\bar{\ell}_i \phi N_\alpha$. The reason
is that this term only contributes for diagrams  whose UV
divergences produce a pole $1/(n-4)$ to cancel the $(n-4)$ pre-factor.
Since $S^\times_\alpha(x,x')=M_\alpha/(p^2 - M_\alpha^2)$ is more
strongly UV convergent than $S_\alpha(p)=\slashed{p}/(p^2 - M_\alpha^2)$, the
two loop diagrams involving the first kind of propagator contain very
few UV divergences. In fact, the vertex correction diagram is UV
finite, with degree of divergence $D=-1$, whilst the propagator
correction diagram contains a single pole $1/(n-4)$, arising from the
propagator correction sub-diagram, which is removed by subtracting an
appropriate counterterm during renormalisation.

\begin{figure}
\centering
\includegraphics[scale=0.37]{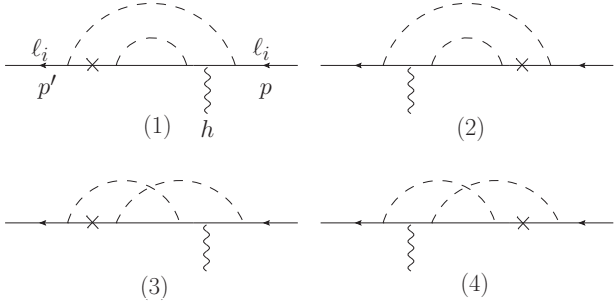}
\caption{Contributions to $\bra{\,\ell(p') \,}  \mathcal{O} \ket{\, \ell(p)\,}$ from the heavy neutrino mass term $\frac{1}{2}h\bar{N} M N^c$. The cross in \textit{e.g.}, $(1)$ denotes the $S^\times_\alpha$ sterile neutrino propagator, and at the $h$ vertex, there are contributions $S_\beta S_\beta$ and $S_\beta^\times S_\beta^\times$  corresponding to each propagator type.}
\label{2loops}
\end{figure}

\begin{figure}
\centering
\includegraphics[scale=0.47]{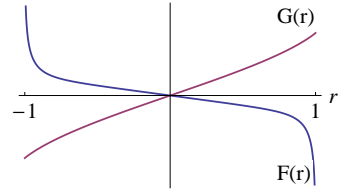}
\caption{The antisymmetric functions contributing to $I_{[\alpha \beta]}$, with $- 1 \leq r \leq 1$.}\label{FG}
\end{figure}

The only remaining terms in (\ref{Omega}) which contribute are the heavy neutrino mass term, and the $\phi^\dagger \phi$ Higgs interactions. 
A full discussion of the these effective Lagrangian calculations will be
presented elsewhere \cite{toappear}. Here, we focus on the
contributions to $\bra{\,\ell(p') \,}  \mathcal{O} \ket{\, \ell(p)\,}$ from the heavy neutrino
couplings to $h$ shown in figure \ref{2loops}. 
 The contribution from these diagrams to the $i q^2 \slashed{q}$ term is:
\begin{equation}\label{pOp}
\bra{\,\ell_i(p') \,}  \mathcal{O} \ket{\, \ell_i(p)\,} = i q^2 \slashed{q} \, h(q)\sum_{\alpha,  \,  \beta, \,
  j} \frac{\mbox{Im}\left[ \lambda^\dagger_{\beta i} \lambda_{i\alpha}
    \lambda^\dagger_{\beta j} \lambda_{j \alpha} \right] }{M_\alpha
  M_\beta}I_{[\alpha \beta]} \ ,
\end{equation}
where $I_{\alpha \beta} = I(M_\alpha,M_\beta)$, and $i$ labels the lepton generation. Note that $I_{\alpha \beta}$ must have a non-vanishing antisymmetric part for (\ref{pOp}) to be non-zero. The contributions to $I$ from each diagram are rather
involved and the complete set of results will be given in
\cite{toappear}.
As an illustration, we quote here the result from diagram (1) to demonstrate explicitly the appearance of a non-vanishing 
contribution to $I_{[\alpha \beta]}$. We find
\begin{align}
&I^{(1)}_{[\alpha \beta]}=  F\left( r\right) + G\left(r \right) \ln\left[\frac{\mu}{M_\alpha + M_\beta}\right] 
\end{align}
where $\mu$ is the mass scale of dimensional regularisation, $r =(M_\alpha - M_\beta)/(M_\alpha+M_\beta)$ and 

\begin{align}
&F(r) = \frac{1}{384 (4\pi)^4  r^4}\Bigg[12 r \left(2 r^2-1\right)-3 \left(r^2-1\right)^2 \ln ^2 \left( \frac{1-r}{1+r}\right)\nonumber \\
& -2\left(2 r \left(5 r^2-3\right)  -3 \left(r^2-1\right)^2 \ln\left(\frac{1-r}{1+r}\right)\right)  \ln \left(\frac{1-r}{2} \right)\nonumber \\
&-2\left(4 r^4-5 r^3-7r^2+3 r+3\right) \ln\left(\frac{1-r}{1+r}\right) \Bigg] \nonumber \\ 
&G(r)= \frac{1}{192 (4\pi)^4 r^4} \Bigg[ 2 r \left(5 r^2-3\right)-3 \left(r^2-1\right)^2 \ln \left(\frac{1-r}{1+r}\right) \Bigg] \nonumber \\
\end{align}
Antisymmetry under interchange of $M_\alpha$ and $M_\beta$ is now manifest from the anti-symmetry of $F(r)$ and $G(r)$ under $r \rightarrow -r$ shown in figure \ref{FG}. 

We have, therefore, shown by explicit calculation that the operator
$\partial_\mu  R \bar{\ell}_i \gamma^\mu \ell_i $ 
is indeed generated, for each lepton flavour with the effective interaction being given by comparing (\ref{pOp}) with figure \ref{effective}:
\begin{equation}\label{dmuR}
\mathcal{L}_i = \partial_\mu R \left( \bar{\ell}_i \gamma^\mu \ell_i
\right) \, \sum_{\alpha,  \,  \beta, \,   j} \frac{\mbox{Im}\left[
    \lambda^\dagger_{\beta i} \lambda_{i\alpha} \lambda^\dagger_{\beta
      j} \lambda_{j \alpha} \right]}{3 M_\alpha M_\beta} I_{[\alpha
  \beta]} \ .
\end{equation}
This demonstrates that a combination of background curvature, complex
couplings ({\it i.e.} C and CP violation) and loop effects can
generate a leptogenesis-inducing operator. The dependence of (\ref{dmuR}) on the non-degeneracy of sterile neutrino masses is discussed in figure 6. 
\begin{figure}
\centering
\includegraphics[scale=0.6,trim=0.3 0.2cm 0.5cm 2cm]{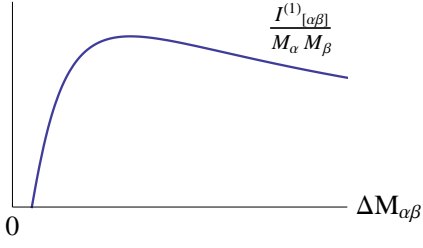}
\caption{For fixed $M_\beta <M_\alpha$, C and CP violation from $\mbox{Im}\left[\lambda^\dagger_{\beta i} \lambda_{i\alpha} \lambda^\dagger_{\beta j} \lambda_{j \alpha} \right]$ is initially enhanced as the mass difference $\Delta M_{\alpha\beta} = M_\alpha - M_\beta$ increases from zero. It then reaches a maximum before tending to zero, as radiative effects become mass-suppressed when $M_\alpha \rightarrow \infty$.}
\end{figure}

\section{Consequences for leptogenesis}
We now describe how this radiatively induced operator leads to a
mechanism of leptogenesis and why the other class of diagrams, with
charge-conserving heavy neutrino propagators, do not. In isotropic
spacetimes, the interaction (\ref{dmuR}) has the form of a chemical
potential $\mu_i$ 
between matter and antimatter for each lepton generation given by
\begin{equation}
\mu_i = \dot{R} \sum_{\alpha,  \,  \beta, \,   j}
\frac{\mbox{Im}\left[ \lambda^\dagger_{\beta i} \lambda_{i\alpha}
    \lambda^\dagger_{\beta j} \lambda_{j \alpha} \right]}{3M_\alpha
  M_\beta} I_{[\alpha \beta]} \ .
\label{ChemPot}
\end{equation}
If $T$ is the temperature of the early universe, this creates a lepton asymmetry of the form 
\begin{equation}
n(\ell_i) -n(\ell_i^c)  =\dot{R}\, T^2  \sum_{\alpha,  \,  \beta, \,
  j} \frac{\mbox{Im}\left[ \lambda^\dagger_{\beta i} \lambda_{i\alpha}
    \lambda^\dagger_{\beta j} \lambda_{j \alpha} \right]}{3M_\alpha
  M_\beta} I_{[\alpha \beta]} \ .
\end{equation}
Summing over all lepton generations, the total lepton asymmetry
($L=\sum_i \ell_i$) is given by
\begin{equation}\label{Lasym}
n(L) - n(L^c) =\dot{R}\, T^2 \, \sum_{\alpha, \,\beta}
\frac{\mbox{Im}\left[ \left(\lambda^\dagger \lambda \right)_{\alpha
      \beta} ^2  \right]}{3 M_\alpha M_\beta}I_{[\alpha \beta]} \ .
\end{equation}
The formula (\ref{Lasym}) is the centrepiece of this Letter.  It
captures how three effects conspire to generate matter-antimatter
asymmetry: the breaking of (time) translational symmetry by gravity in
$\dot{R}$, C and CP violation from
$\mbox{Im}\left[\left(\lambda^\dagger \lambda \right)_{\alpha \beta}
  ^2  \right]$ and quantum loop effects in $I_{\alpha \beta}$. In
particular, this mechanism remains active at energies and temperatures
below the heavy scale and so is able to generate an asymmetry
\textit{after} the heavy neutrino decays, where the asymmetry is
maintained in equilibrium by the $\Delta L=2$ reactions $\phi \ell^c \leftrightarrow \phi^* \ell$.

Now that we have revealed the bigger picture, we are able to explain why the 2 loop contributions involving the charge-conserving propagators $\braket{N(x) \overline{N}(x')}$ are of less interest for leptogenesis. If we had instead calculated contributions from diagrams with this type of propagator, we would have found a different Yukawa matrix structure in the amplitude, leading to a generational lepton asymmetry
\begin{equation}
n(\ell_i) -n(\ell_i^c)  = \dot{R}\, T^2  \sum_{\alpha,  \,  \beta, \,
  j} \mbox{Im}\left[ \lambda^\dagger_{\beta i} \lambda_{i\alpha}
  \lambda^\dagger_{\alpha j} \lambda_{j \beta} \right] J_{\, [\alpha\,
  \beta]} \ .
\end{equation}
While this gives an asymmetry for each flavour, summing over all generations gives $n(L) - n(L^c)  \propto \sum_{\alpha,  \,  \beta}
\mbox{Im}\left[ (\lambda^\dagger \lambda )_{\beta \alpha} ( \lambda^\dagger \lambda)_{\alpha \beta}\right] J_{\, [\alpha\, \beta]}$. However, $\mbox{Im}\left[ (\lambda^\dagger  \lambda )_{\beta \alpha}( \lambda^\dagger \lambda)_{\alpha \beta}\right] = \mbox{Im} \left| \left( \lambda^\dagger \lambda\right)_{\beta \alpha} \right|^2 =0$, and so the total lepton asymmetry from these diagrams is zero.

\section{Discussion}

In this Letter, we have presented a new mechanism -- radiatively-induced gravitational leptogenesis -- for generating matter-antimatter
asymmetry. We have shown how leptons and antileptons can propagate
differently in curved spacetime due to gravitational interactions with
their self-energy cloud of virtual high-mass particles. This effect is
forbidden in flat space by CPT and translation invariance, and at
tree-level in curved spacetime, by the strong equivalence principle. At
loop level, however, the strong equivalence principle no longer holds
and, depending on the composition of the cloud,  C and CP violating operators can be generated in the low-energy effective Lagrangian. A simple interpretation in terms of a chemical potential for leptons shows immediately that this generates an asymmetry in the equilibrium distributions of matter and antimatter.

%In a longer companion paper \cite{toappear}, we will present these ideas in greater depth, providing more detail of the explicit calculations quoted above together with a more complete discussion of the theoretical basis of the generation of matter-antimatter asymmetry than the simple chemical potential argument given here.

As already noted, this mechanism is very general, and its implementation in the specific Fukugita-Yanagida model described here is just one example.
In particular, it arises naturally in most existing models of leptogenesis, which typically involve a high-energy BSM sector with C and CP violation,
where it generates a matter-antimatter asymmetry at low energies and temperatures after the decay and decoupling of the heavy particles.

%Moreover, the specific parameter dependence of the chemical potential
%we found in Eq.(\ref{Lasym}) shows that our mechanism will
%produce asymmetries of the same general order as existing leptogenesis models.

The next step is therefore to implement this mechanism within specific phenomenologies, {\it e.g.} GUT, SUSY and other leptogenesis
models, giving a more thorough analysis of kinetic aspects of these theories. This would involve a discussion of Boltzmann equations, decoupling temperatures, reaction rates and the strength of curvature at various times in the Universe's history, {\it e.g.} inflation, radiation, matter domination. For instance, work is currently under way \cite{toappearJ} to study the present leptogenesis model in warm inflation, where both temperature and curvature are high. Such analyses will allow us to see in what situations this mechanism can quantitatively account for the observed matter-antimatter asymmetry in the Universe. 
\\
\section*{Acknowledgements}
JIM would like to thank Tim Hollowood for useful conversations which prompted the detailed analysis of propagation in section 2. GMS is grateful to the Theory Division, CERN for hospitality.  This research is funded in part by STFC grants ST/K502376/1 and ST/L000369/1.
\\

%In this Letter, we have given a qualitative description of a mechanism for radiatively induced gravitational leptogenesis, illustrating it within a %familiar model. The next chapter in the story is to apply this general scenario to specific phenomenologies, e.g. GUT, SUSY and other leptogenesis
% models, at specific times in the Universe's history, e.g. inflation, radiation domination, matter domination etc so as to investigate the quantitative %question of whether such gravitational mechanism can account for the measured baryon asymmetry.

%This involves a more detailed discussion of the kinetic aspects, including reaction rates, decoupling temperatures and experimental constrains on %the parameters of the theory. This has already been studied in the context of gravitational baryogenesis by the authors cited in the opening %paragraph of this letter and can successfully account for the observed baryon to entropy ratio of the universe. 

%
%\section*{}
%\noindent Still need to add:
%\begin{enumerate}
%\item Why haven't we calculated the $\phi$ contributions.
%\item A better explanation of the linear graviton method. 
%\item How do we word the fact we haven't calculated the hard vertex correction?
%\end{enumerate}

\end{document}